\documentclass[sigconf]{acmart}
\acmConference[ISSTA 2023]{ACM SIGSOFT International Symposium on Software Testing and Analysis}{17-21 July, 2023}{Seattle, USA}

\usepackage{amsmath,amssymb,amsfonts}
\usepackage[most]{tcolorbox}
\usepackage{algorithmic}
\usepackage{graphicx}
\usepackage{textcomp}
\usepackage{xcolor}
\usepackage{listings}
\usepackage{xspace}
\usepackage{subfigure}
\usepackage{cleveref}

\newcommand{\name}{{BAPP}\xspace}
\def\BibTeX{{\rm B\kern-.05em{\sc i\kern-.025em b}\kern-.08em
    T\kern-.1667em\lower.7ex\hbox{E}\kern-.125emX}}
\begin{document}

\title{A Bayesian Framework for Automated Debugging}

\author{Sungmin Kang}
\authornote{Both authors contributed equally to this research.}
\affiliation{%
    \institution{KAIST}
    \city{Daejeon}
    \country{South Korea}
}

\author{Wonkeun Choi}
\authornotemark[1]
\affiliation{%
    \institution{KAIST}
    \city{Daejeon}
    \country{South Korea}
}

\author{Shin Yoo}
\affiliation{%
    \institution{KAIST}
    \city{Daejeon}
    \country{South Korea}
}




\begin{abstract}
Debugging takes up a significant portion of developer time. As a result, 
automated debugging techniques including Fault Localization (FL) and Automated 
Program Repair (APR) have garnered significant attention due to their potential 
to aid developers in debugging tasks. Despite intensive research on these 
subjects, we are unaware of a theoretic framework that highlights the 
principles behind automated debugging and allows abstract analysis of 
techniques. Such a framework would heighten our understanding of the endeavor 
and provide a way to formally analyze techniques and approaches. To this end, 
we first propose a Bayesian framework of understanding automated repair and 
find that in conjunction with a concrete statement of the objective of 
automated debugging, we can recover maximal fault localization formulae from 
prior work, as well as analyze existing APR techniques and their underlying 
assumptions.

As a means of empirically demonstrating our framework, we further propose 
\name, a Bayesian Patch Prioritization technique that incorporates intermediate program values 
to analyze likely patch locations and repair actions, with its core equations 
being derived by our Bayesian framework. We find that incorporating program 
values allows \name to identify correct patches more precisely: when applied to 
the patches generated by kPAR, the rankings produced by \name reduce the number of 
required patch validation by 68\% and consequently reduce the repair 
time by 34 minutes on average. Further, \name improves the precision of FL, 
increasing acc@5 on the studied bugs from 8 to 11. 
These results highlight the potential of value-cognizant automated debugging 
techniques, and further validates our theoretical framework. Finally, 
future directions that the framework suggests are provided.
\end{abstract}

\maketitle


\section{Introduction}


Debugging plays a crucial role in the development process, as it is difficult 
to write a correct program on the first attempt, particularly when the program 
is large. As a result, significant resources are spent on debugging: Tassey 
notes that manual debugging can be tedious and demanding~\cite{debugmotiv2022}. 
To aid developers in the debugging process, automated debugging tasks such as 
Fault Localization (FL)~\cite{Jones:2002kx} or Automated Program Repair 
(APR)~\cite{LeGous2012GenProg} were proposed to reduce developer burden when 
debugging issues. These techniques have matured enough to be applied in 
corporations and actively help developers \cite{Kirbas2021BloombergAPR, 
Marginean2019SapFix}.

As the body of work on automated debugging steadily grows, the importance of 
understanding the underpinning principles of the field is also increasing. 
While other software engineering fields have seen theories 
proposed~\cite{Goodenough1975TT} that prompted discussions~\cite{Weyuker1980RS},
we are unaware of an attempt to provide a framework for understanding automated
debugging as a whole. Nonetheless, having a formal framework to analyze 
automated debugging techniques would not only help understanding techniques published up to now and their empirical results, but may also suggest nuances that could be easy to miss without a formal treatment, and additionally provide interesting directions for future research.

To this end, we first suggest a Bayesian framework of automated debugging. We 
posit that the purpose of automated debugging techniques is ultimately to 
jointly infer the likely location and repair action for a fix, and suggest 
that, by using Bayes' theorem, our framework can map probabilistic terms to 
well-known automated debugging concepts such as FL or APR. To firmly establish 
how Bayes' theorem can be useful in analyzing automated debugging techniques, 
we first derive SBFL formulae based on a minimal set of assumptions
and find that the resulting SBFL formulae are equivalent to the maximal SBFL 
formulae as proven by Yoo et al.~\cite{Yoo:2014fv}. In addition, we analyze the 
behavior of well-known APR techniques, and find that they can be expressed 
within our Bayesian framework. In this process, we identify a practical idea 
regarding the use of FL in APR that we validate through experiments.

As a way of empirically testing our framework, we seek to tackle an important 
problem for generate-and-validate APR techniques that generate a large number 
of patches. As template-based APR techniques incorporate more and more 
templates, the number of patches they need to validate is also 
increasing~\cite{Koyuncu2020FixMiner}, leading to long execution times. As a 
result, precise patch ranking techniques are required to quickly find the 
correct patch from a large template space. Existing work shows that humans 
generally find fixes with significantly fewer validation attempts, and heavily 
use program values during the process~\cite{Ceccato2012HumanAPR}. This 
naturally leads to the question: \emph{can we use program values to efficiently 
identify fixing patches for APR techniques?} While some techniques do indeed 
use program values, they are almost exclusively focused on the 
generation of if statements~\cite{Mechtaev2016rw, Chen2018JAID}; we hope to 
propose a more general approach.

To this end, we propose \name (Bayesian Automated Patch Prioritization), a patch prioritization technique that incorporates program 
values. The core intuition is that patches must lead to program
behavior change (i.e., a change in variable value or control flow) in failing 
tests while it is unlikely yet possible that they lead to behavior change in 
passing tests. We utilize our theoretic framework to derive precise 
expressions that are used by \name to rank patches. While the approach is sound 
in terms of the framework, efficiently extracting the program behavior is a 
non-trivial technical challenge: we solve this problem via the use of 
\emph{debuggers}, which allow 
compilation-free lightweight evaluation of expressions necessary to gauge 
program behavior change. The use of debuggers allows us to evaluate thousands 
of patches within minutes, allowing our theory to benefit APR in practice.
Starting with the fix operators from kPAR~\cite{Liu2019kpar}, we empirically 
evaluate whether our approach can identify patches that pass regression tests 
efficiently.


Our results indicate that the incorporation of program states in ranking the 
patches successfully increased the efficiency from the original kPAR approach, 
with the median reduction of the plausible patch rank measured at 0.68. 
Execution time also saw a significant improvement, 
resulting in an average reduction of 34 minutes. A new scheme 
for generating an FL ranking of suspicious locations with the consideration of 
program states was also shown to perform better than SBFL in finding the true 
buggy line; \name improved the identification of the buggy line within five 
attempts (acc@5) from SBFL's 8 to 11. Finally, it was 
found that a slightly higher weight to the score from program states than to the 
score from SBFL was best for identifying the plausible patch, indicating the 
incorporation of program state information was useful.


With our theory verified as such, we propose future research directions that we 
could identify via the theory; for example, based on a different decomposition 
of probabilistic terms in our framework, we could identify a novel way of 
approaching automated debugging (specifically, instead of performing FL first, 
identifying the type of patch first). Further, we present limitations of our 
theory on which we hope to perform further research.


Overall, our contributions are:

\begin{itemize}
    \item A Bayesian framework that explains a number of known automated debugging results and practices;
    \item A patch prioritization technique, \name, which we build upon this Bayesian framework, and
    \item Extensive empirical experiments demonstrating how \name can significantly improve APR efficiency.
\end{itemize}

The organization of this paper is as follows. We first present related work in \Cref{sec:relwork}.
We present our framework and its relationship
to existing automated debugging literature in Section~\ref{sec:framework}.
Our approach is outlined in \Cref{sec:approach}, while our evaluation setup is
described in \Cref{sec:setup}. Based on this, the results of our experiments are provided
in \Cref{sec:results}. 
We discuss future work and threats to validity in \Cref{sec:discussion}, and \Cref{sec:conclusion} concludes.

\section{Related Work}
\label{sec:relwork}

This section provides academic context for our work.

\subsection{Theories for Automated Debugging}
\label{sec:relwork_theory}

Most of existing work on automated debugging focus on designing techniques 
that are effective; relatively little has been done to examine the theoretical 
aspect of automated debugging. Weimer et al. observed the duality between APR 
and mutation testing~\cite{Weimer2013ma}, which can be thought to have provided 
the foundations for the subsequent work on Mutation Based Fault 
Localization (MBFL)~\cite{Moon2014MUSE,Lou2020rw}. Xie et al.~\cite{Xie:2013uq} 
proposed a theoretical framework for proving hierarchy between Spectrum Based 
Fault Localization (SBFL) formulas, which eventually resulted in the no 
existence proof for the greatest formula (i.e., there is no single formula that 
is guaranteed to outperform all the other formulas)~\cite{Yoo:2017ss}, 
prompting the FL research community to focus on the aggregated use of multiple 
formulas and extra input features~\cite{Li2019aa}, rather than designing new 
single formula. However, both of the existing theoretical results on APR and FL are 
limited to the underlying formulations, i.e., mutation and spectrum-based approaches, respectively. We still lack a general framework that can express MBFL and SBFL, as well as various APR techniques, and more. We believe that such a general framework may allow us to rigorously reflect on existing techniques and propose new and interesting future research directions.


\subsection{Automated Debugging}

Automated debugging has a long history~\cite{Jones:2002kx} and thus it is 
difficult to summarize all techniques in the scope of this paper. In this work 
we focus on test-based automated debugging techniques, which we find ideal when 
applying the Bayesian inference toolkit. Test-based automated debugging 
techniques can be roughly categorized into FL and APR. Test-based FL generally 
seeks to identify the part of a project that needs to be fixed given a number 
of failing tests and potentially passing tests. Researchers have identified 
multiple ways to do this, including the use of program 
spectrum~\cite{Jones:2002kx}, mutation testing~\cite{Moon2014MUSE}, project 
history~\cite{Wen2021HSBFL}, and more. Test-based APR seeks to change the 
source code of a project so that all tests pass, and ideally so that the patch 
is semantically equivalent to the patch that the developer would have made. 
As with FL, there are multiple approaches: while the first APR technique,
GenProg~\cite{LeGous2012GenProg}, used genetic algorithms to create patches,
other ways to generate patches subsequently emerged, such as using templates
as with PAR~\cite{Kim:2013ty}, generating constraints that patches should meet
then solving those constraints with SMT solvers, as with Angelix~\cite{Mechtaev2016rw},
or by using deep neural networks~\cite{Chen2019SequenceRSL}.

\subsection{Program Values in Automated Debugging}

A large number of automated debugging techniques do not explicitly use concrete 
program values in any way; for example, most FL techniques do not explicitly 
use program values~\cite{Jones:2002kx, Moon2014MUSE, Li2019aa}, and a 
significant number of APR techniques focus on generating the correct patch 
given the static context rather than incorporating values. Nonetheless, there 
have been attempts to incorporate values into the automated debugging process, 
as humans do~\cite{Ceccato2012HumanAPR}. SmartFL~\cite{Zeng2022SmartFL} 
generates a detailed probabilistic graph of a program that incorporates values 
into its inference process, but due to the potentially large graphs that are 
generated, inference can be slow. For APR, Angelix~\cite{Mechtaev2016rw} 
identifies angelic values that allow a test to pass, while 
Dynamoth~\cite{Durieux2016dynamoth} used a debugger to similarly check if 
certain predicates met angelic value conditions. As our theory provides a 
holistic view of APR and FL, our tool paves a way to consider values and tackle 
the automated debugging problem as a whole.

\subsection{Patch Prioritization}
\label{sec:relwork_prioritization}

As generate-and-validate APR techniques improved and increased their search 
space, the importance of patch prioritization has also grown, and multiple 
techniques have been suggested; as it is difficult to give a full overview of 
all techniques within this paper we introduce a cross-section of explored 
approaches. To improve fault localization during the patch validation process, 
Unified Debugging~\cite{Lou2020dt} techniques have been proposed to improve the 
precision of FL while doing patch validation. Meanwhile, some techniques seek 
to optimize the patch template to apply: for example, 
Prophet~\cite{Long2016Prophet} mines statistics of patches to precisely 
apply templates. Other techniques prioritize patches based on the specific code 
snippet they introduce: ELIXIR~\cite{Saha2017Elixir} uses manually constructed 
features to identify patch ingredients to be used when applying a patch 
template. Our prioritization approach differs as it in using program values as a 
means of calculating patch ranking, and as a result is orthogonal with the 
aforementioned techniques.

\section{Framework}
\label{sec:framework}

We seek to present a unified framework for automated debugging techniques 
described in the prior section.

\subsection{Bayesian Inference}

Bayesian inference is a way of updating probabilities or beliefs in response to 
new information, based on Bayes' theorem. In particular, given evidence or 
observations $E$, a hypothesis related to the evidence $H$, and the prior 
belief in the hypothesis $P(H)$, Bayesian inference postulates that the 
probability of a hypothesis given evidence, $P(H|E)$, can be calculated as the 
following:
\begin{equation*}
    P(H|E) = \frac{P(E|H)P(H)}{P(E)}
\end{equation*}

In Bayesian terminology, $P(H)$ is the \emph{prior probability}; in contrast, $P(H|E)$ is the \emph{posterior probability}, which is the updated belief after observing evidence $E$. To calculate the posterior, one needs a \emph{statistical model} that can determine the probability of the evidence assuming that the hypothesis is correct, $P(E|H)$. The $P(E)$ term is a normalization term that does not influence the ranking of results, and thus may be ignored for our purposes.

Priors may be iteratively applied; in the face of new evidence $E'$, the
probability of the hypothesis given both pieces of information (assuming the two are uncorrelated)
, $P(H|E, E')$, is given as
\begin{equation*}
    P(H|E, E')=\frac{P(E'|H)P(H|E)}{P(E')}
\end{equation*}

\noindent showing that the previous posterior can be used as the prior when 
inferring given new evidence $E'$.

\subsection{Application on Automated Debugging}

We argue that the primary goal of automated debugging techniques is to find the
likely fault location $l$ and the appropriate fix action $a$. Stated in 
probabilistic terms, the objective of automated debugging techniques is to 
infer the values of $P(l, a)$.\footnote{We use $P(l, a)$ as a shorthand for 
$P(l=\text{fault location} \wedge a=\text{fix action})$ throughout the paper.}
Test-based automated debugging techniques may use dynamic information $D$,
such as the results of individual tests, test suites, or (as we later do) 
program values to precisely infer the value of $P(l, a)$. Overall, we can say 
automated debugging techniques aim to infer $P(l, a|D)$, or the probability of 
certain patches given data $D$. As a result, we argue that the test-based 
automated debugging scenario can be effectively modeled using Bayesian 
inference, as the formula below denotes:
\begin{equation}
    P(l, a|D) \propto P(D|l, a)P(l, a)
    \label{eq:ad_theory}
\end{equation}

\noindent The formula above is in fact an application of Bayes' theorem without the denominator term,
as the denominator is a normalization term that is the same for every patch $(l, a)$, and thus
has no effect on the relative ranking between patches.
The formula can be used to understand automated debugging techniques in various 
ways. For example, $P(l, a)$ can be decomposed to $P(l, a)=P(a|l)P(l)$; this 
can be thought to represent the separation of APR ($P(a|l)$) and FL ($P(l)$) 
techniques, as we describe in later sections. Additionally, we find that 
different families of automated debugging techniques differ in how they model 
the calculation of $P(D|l, a)$; concretely stating their models provides a 
useful window to inspect and compare techniques.

Finally, let us now turn to how fault localization fits in this model of 
automated debugging. We argue that FL is a special case of automated debugging, 
for if one marginalizes \Cref{eq:ad_theory} over actions $a$, we end up with:
\begin{equation}
    P(l|D) \propto P(D|l)P(l)
    \label{eq:fl_theory}
\end{equation}

\noindent which can be used to derive maximal SBFL formulae, as demonstrated in 
the next subsection.

\subsection{Fault Localization}

As a demonstration of our framework, we construct a statistical model based on 
the assumptions of prior theoretic work on FL~\cite{Yoo:2014fv} and show that, 
in conjunction with the Bayesian inference formula for fault localization 
(\Cref{eq:fl_theory}), we can recover formulae that were proven to be maximal, 
i.e. as close to optimal as an SBFL formula can get. Specifically,
Yoo et al. analyze spectrum-based fault localization techniques, which 
use \textit{program spectrum}. Program spectrum is a set of numbers that 
characterize how the test suite of the program interacts with each program 
element; in the paper, they notate spectrum with $e_f, e_p, n_f, n_p$ which 
denote the number of \textbf{f}ailing tests that \textbf{e}xecuted a location, 
the number of \textbf{p}assing tests that executed a location, the number of 
failing tests that did \textbf{n}ot execute a location, and the number of 
passing tests that did not execute a location, respectively. Further, the total 
number of failing tests is denoted as $F$.

Yoo et al.~\cite{Yoo:2014fv} make three assumptions about bugs in their analysis:
(i) that there is a single fault in the code, (ii) that the code is 
deterministic, and (iii) that there is at least one failing test case. Upon 
these assumptions, we build the following statistical model that provides the 
probability a test will fail given coverage information and the true fault 
location:
\begin{align*}
    P(t=\text{fail}|l=\text{fault} \wedge l \in_{c} t) = p \\
    P(t=\text{fail}|l=\text{fault} \wedge l \notin_{c} t) = 0
\end{align*}

\noindent The first equation is simply stating that `if the true fault location $l$ is covered by a test $t$,
the probability that $t$ will fail is a nonzero $p$.' The second equation states that
`if the true fault location $l$ is \textit{not} covered by $t$, it will never fail'.
This model naturally follows from the previously stated assumptions.

Before we proceed further, we must set a prior $P(l)$ probability of each location being the
true fault location. For simplicity we use the uniform prior: that is, all lines are equally
suspicious when there is no information. Specifically, given the full set of statements $L$,
$P(l)=\frac{1}{|L|}$.
We note that one may opt to use different priors,
such as differentiating based on statement type, to more closely represent the actual bug
distribution, which is non-uniform~\cite{Pan2009BFP}.

With the prior and statistical model determined, we may now perform Bayesian inference.
Suppose we observe a test $t$ that fails and does not cover $l$; how likely is it
that $l$ is the true bug location? Bayesian inference asks the reverse question:
assuming that $l$ is the true bug location, how likely is it that $t$ fails? Then,
it combines this with the prior to answer our original question, the probability
$l$ actually is the true bug location given that $t$ has failed.
\begin{align*}
    P(l|t=\text{fail} \wedge l \notin_{c} t) &\propto P(t=\text{fail}|l=\text{fault} \wedge l \notin_{c} t)P(l) \\
     &= 0 \times \frac{1}{|L|} = 0
\end{align*}

\noindent Thus, through Bayesian inference, we can deduce that that locations 
not covered by the failing test $t$ cannot be related to the bug.
Similar principles can be applied to the other test scenarios as well;
thus, given the first test, we update the probability that each location is the 
true fault location as:
\begin{equation*}
    P(l|t) \propto \begin{cases}
        p & (t=\text{fail} \wedge l \in_{c} t) \\
        1-p & (t=\text{pass} \wedge l \in_{c} t) \\
        1 & (t=\text{fail} \wedge l \notin_{c} t) \\
        0 & (t=\text{pass} \wedge l \notin_{c} t)
    \end{cases}
\end{equation*}

\noindent with the prior dropped because it is the same at every location. 
Using the fact that previous posteriors can be used as new priors, and that the 
four cases above neatly map to the $e_f, e_p, n_f, n_p$ spectra described 
earlier, we can iteratively derive the posterior probability that a location is 
a fault given the entire test suite:
\begin{equation*}
    P(l=\text{fault}|t_1, ..., t_n) \propto 0^{n_f}1^{n_p}p^{e_f}(1-p)^{e_p}
    \label{eq:middle_naish1}
\end{equation*}

\noindent While the formula above seems to have four variables, in terms of 
determining the ranking the formula can be further simplified. First, if 
$n_f \neq 0$, $P(l=\text{fault}|t_1, ..., t_n)=0$, so the other factors are 
unimportant. For all statements for which $n_f=0$, $e_f=F$ holds as well, so is 
irrelevant in terms of ranking; $1^{n_p}=1$ as well, making $e_p$ the only 
deciding factor in determining suspiciousness. As a result, we get the 
following:
\begin{equation}
    P(l=\text{fault}|t_1, ..., t_n) \propto \begin{cases}
        0 & e_f < F \\
        (1-p)^{e_p} & e_f = F
    \end{cases}
    \label{eq:middle_naish1}
\end{equation}

As long as $0 < p < 1$, this leads to the same rankings as the Naish01 SBFL formula
identified to be one of the maximal formulae~\cite{Yoo:2014fv}. Further, in the
limit when $p \rightarrow 0$, this becomes equivalent to another maximal formula, Binary. 
Thus, using our 
framework, one can quickly arrive at maximal equations under a given set of 
assumptions. This also allows one to assess how good a statistical model 
describes real behavior of code; for example, if the Binary SBFL formula shows good
performance, it would mean that $p \ll 1$, and thus there may be many tests that
are passing due to coincidental correctness.

We close by making a few observations. First, while we make the single-fault 
assumption as it greatly simplifies matters, one can perform inference for 
multiple faults as well, as done by Barinel~\cite{Abreu:2009qy}. 
In addition, our framework can be used to
derive other fault localization formulae; for example, our appendix provides a 
detailed derivation of an MBFL formula similar to MUSE~\cite{Moon2014MUSE}.

\subsection{Automated Program Repair}

APR techniques can be analyzed using our framework as well. In our paper, we 
use the taxonomy of APR techniques proposed by Le Goues et 
al.~\cite{LG2021APR}, which divides APR techniques into two groups: 
heuristic-based and constraint-based. While there is growing interest in APR 
techniques that employ deep learning~\cite{Chen2019SequenceRSL}, their 
principles are largely similar to other heuristics-based techniques, as we 
explore through this subsection.

\emph{Heuristic-based}, or Generate and Validate (G\&V), APR techniques 
generally first take the list of suspicious statements provided by a fault 
localization technique, then use heuristics to generate a number of patches at 
each location. Each patch is then evaluated against the tests that are present 
in the project: if a patch makes all tests pass, the patch is deemed \textit{
plausible} and becomes a candidate for suggestion to the developer. This 
process is naturally captured by the Bayesian formulation of automated 
debugging: along with the decomposition $P(l, a)=P(a|l)P(l)$, we may infer the 
posterior probability of $P(l, a)$ as
\begin{equation}
    P(l, a|D_{(l, a)}) \propto P(D_{(l, a)}|l, a)P(a|l)P(l)
\end{equation}

\noindent where $D_{(l, a)}$ term represents the test execution results after 
applying repair action $a$ on location $l$, while the $P(a|l)$ and $P(l)$ terms 
represent the patch generation heuristics and the fault localization processes, 
respectively. What, then, is the statistical model being used to update patch 
probabilities? We find that the validation process of G\&V techniques is 
well-expressed by a simple conditional probability model. If we denote that a test $t$ passed under patch $(l, a)$ as $t_{(l, a)}=\text{pass}$, we can set the following statistical model which replicates the validation process:
\begin{align}
    P(\forall t.t_{(l, a)}=\text{pass} |(l, a) = \text{fix}) = 1 
    \label{eq:throw_nonplausibles}\\
    P(\forall t.t_{(l', a')}=\text{pass}|(l, a) = \text{fix}) = p
\end{align}

\noindent where $(l, a) \neq (l', a')$. That is, if the patch is the true fix, 
it should make all tests pass; the second row indicates the possibility that a 
different patch may also lead to all tests passing. Expanding Bayesian rules as 
we did in the previous subsection leads to the usual validation criterion that 
tries each patch one by one and discards those that cause test failures. 
Meanwhile, we note that the statistical model above has no special cases when 
the patches are related, e.g. when $l = l'$; one may say that the addition of 
such special cases is what characterizes the unified debugging techniques, as 
explained later in this section.

If the statistical model is this simple, what are G\&V techniques improving? 
Particularly with the advent of deep learning-based techniques, we may say that 
latest APR techniques are improving the \textit{prior distribution} of 
patches, in particular the $P(a|l)$ term that describes which repair actions 
are likely given a specific location. While this probability is 
implicit in techniques such as template-based APR, in deep learning-based APR 
techniques the probabilistic nature is explicit, as the neural models will 
generate probabilities for each of the patches that they generate. While neural 
APR techniques are showing improvements each year~\cite{Zhu2021Recoder}, this 
analysis shows that they rely on the same dynamic update model as earlier
APR work~\cite{Kim:2013ty}.

Meanwhile, one interesting suggestion that our framework makes that differs 
from usual practice is that it recommends \emph{multiplying} the patch 
probability at a location and the suspiciousness of a location, instead of 
having fault localization results prioritized over patch likelihood as is usual 
practice~\cite{Zhu2021Recoder}. As we observe in our results, this small 
tweak leads to a significant performance boost in our tool.

\emph{Constraint-based} APR techniques often rely on constructing \textit{constraints} that patches
should satisfy in order to fix the patch.
Many techniques use SMT solvers to solve these constraints; as a result,
they rely less on having strong prior distributions $P(l, a)$. For example,
Angelix~\cite{Mechtaev2016rw} uses SBFL results and has a less restrictive $P(a|l)$,
while DirectFix~\cite{Mechtaev:2015aa} does not use external FL results at all, essentially using
a uniform prior $P(l)$.

We analyze Angelix as an example to show how constraint-based techniques can be understood
under our framework. 
To simplify the operations of Angelix, for each test $t$ an angelic value $v_t$ is derived
for fix expression $a$ at a location $l$; the values are `angelic' because
if the value of the expression $a$ at $t$ becomes equivalent to $v_t$, the test will pass.
For example, Angelix might derive that a certain predicate must evaluate to \texttt{true}
for a previously failing test to pass.
For passing tests, $v_t$ is set to maintain the existing behavior, while for failing tests a value
that makes the test pass is found, e.g. using SMT solvers~\cite{Mechtaev2016rw}. The value of the fix expression
when executing test $t$, $[\![a]\!]_t$, is expected to be $v_t$ on all tests,
\begin{equation}
    P([\![a]\!]_t=v_t|(l, a)=\text{fix}) = 1
    \label{eq:constraint_theory}
\end{equation}

\noindent and any patches that deviate from the angelic values at any test are discarded.
Note that there is no distinction between passing and failing tests in \Cref{eq:constraint_theory},
which distinguishes the constraint-based techniques from the update rules of \name introduced in \Cref{sec:approach}.

\subsection{Unified Debugging}

Recently, \emph{unified debugging} has been proposed as a way to integrate the FL and APR process~\cite{Lou2020dt}. 
While there are a number of proposed techniques, we
focus our analysis on the recent SeAPR~\cite{Benton2022SeAPR} technique, as it provides a relatively simple approach in which
our framework can re-derive the core assumptions and make recommendations on the equation form.
Under the automated debugging formulation in ~\Cref{eq:ad_theory}, there might actually be two
ways to integrate the FL and APR process. The first is to infer the prior distribution of patches
$P(l, a)$ without dividing the process into separate steps (e.g. into $P(a|l)P(l)$).
This is not what unified debugging up to now has done; instead, they suggest new ranking update rules
based on dynamic information, thus changing $P(D|(l, a))$.

The SeAPR technique first defines `high-quality patches' as patches that make at least
one previously failing test pass when applied. Based on this, SeAPR assigns higher priority to patches 
that modify the same locations as high-quality patches. Their assumptions can be transformed into
a statistical model under our framework; adding the single fault assumption for simplification we
can formulate the model as
\begin{align}
    P(\exists t. (t=\text{fail}\wedge t_{(l, a')}=\text{pass})|(l, a) = \text{fix}) = p_1 
    \label{eq:seapr_related_patch_update}\\
    P(\exists t. (t=\text{fail}\wedge t_{(l', a')}=\text{pass})|(l, a) = \text{fix}) = p_2
    \label{eq:seapr_unrelated_patch_update}
\end{align}

\noindent where $l \neq l' \wedge p_1 > p_2$. In particular, \Cref{eq:seapr_related_patch_update}
describes the special rule for related patches which was not in the statistical model of usual
G\&V approaches.
Along with the `discard patches that fail tests' criterion 
provided in \Cref{eq:throw_nonplausibles}, 
the statistical model can be used to derive probabilities of each patch being the true fix based on
our framework.
Noteworthy in the statistical model that we build based on the SeAPR settings
is that fail-to-pass tests can appear in patches unrelated
to the true fix (\Cref{eq:seapr_unrelated_patch_update}), 
unlike in the FL model where tests could not fail without covering the true fault location,
leading to a different suspiciousness formulation. In fact, after simplification, we find that
\begin{equation}
    \text{log}(P((l, a)=\text{fix}|D)) \propto p^+ - \gamma p^-
\end{equation}

\noindent where $p^+$ is the number of high-quality patches at $l$, while $p^-$ is the number of low-quality
patches at $l$, and $\gamma=\frac{\text{log}((1-p_2)/(1-p_1))}{\text{log}(p_1/p_2)}$. This is in fact
equivalent to the Wong2~\cite{Wong2007SBFL} SBFL formula when $\alpha=1$. Unfortunately, the SeAPR
publication~\cite{Benton2022SeAPR} did not experiment with the Wong2 formula, so it is unclear to what extent
the empirical results presented in that paper support our framework.

Nonetheless, we believe this analysis demonstrates the utility of our framework. Benton et al.~\cite{Benton2022SeAPR}
argued that the use of APR results can be mapped to coverage spectrum analogues and thus made the
assumption that SBFL formulae may be similarly used in unified debugging.
However, our framework allows an inspection of the
assumptions behind the model, and further shows that the assumptions are different from FL. 
Finally, our framework suggests a formula not
studied in the original work, showing its capability of making practical suggestions
that may not be considered without the use of a theoretic framework. 

\section{Approach}
\label{sec:approach}

So far, we have explored various branches of the automated debugging field and
shown that a multitude of prior results can be understood through our Bayesian 
framework. In this section, we use our framework as the basis to derive a novel 
patch prioritization technique, \name, to incorporate values and efficiently 
identify promising patches.

\subsection{Assumptions and Derivation}
\label{sec:approach_derivation}

To contribute towards solving the important problem of patch prioritization, we 
use our Bayesian framework to derive a formula for our tool, \name. First, we 
construct a statistical model based on the principle of \emph{behavior change}. 
Specifically, we observe that (i) correct patches must alter the behavior of 
failing tests, and (ii) that it is unlikely, yet possible, that they may alter 
the behavior of passing tests. For example, when adding the statement 
\texttt{if (v == null) return;} to a location, the behavior would change if 
there is at least one test execution in which \texttt{v == null}; otherwise the 
patch would not alter the behavior of the test. We note that if a location is 
executed multiple times, it is sufficient for the behavior to be changed at
just one point to alter test behavior. Thus, we can formally specify these 
assumptions into the following statistical model, with $\text{Ch}(t, (l, a))$ 
denoting that patch $(l, a)$ would alter the behavior of test $t$:
\begin{align}
    P(t = \text{failing} | (l, a) = \text{fix} \wedge \text{Ch}(t, (l, a))) = p \\
    P(t = \text{failing} | (l, a) = \text{fix} \wedge \neg \text{Ch}(t, (l, a))) = 0
    \label{eq:taper_cp_assump}
\end{align}

\noindent where $0<p<1$; namely, if a patch changes failing test behavior it has
a chance to be the true patch, while if a patch does not change failing test behavior
there is no chance it is the true patch.

We denote the following `spectrum' to represent program change for a specific 
patch $(l, a)$: $c_f$ denotes the number of failing tests for which $(l, a)$ 
changes behavior; $c_p$ denotes the number of changed passing tests, $n_f$ 
denotes the number of unchanged failing tests, and $n_p$ finally denotes the 
number of unchanged passing tests. Further, we use the decomposition 
$P(l, a)=P(a|l)P(l)$, for which $P(a|l)$ and $P(l)$ may be any patch-generating 
and FL technique, respectively. From this we can derive:
\begin{equation}
    P((l, a)=\text{fix}|D) \propto (0^{n_f}1^{n_p}p^{c_f}(1-p)^{c_p})P(a|l)P(l)
\end{equation}

Handling the $c_f < F$ case separately and removing terms that are unrelated to 
ranking similarly to the SBFL case, we end up with the following formula:
\begin{equation}
    P((l, a)=\text{fix}|D) \propto \begin{cases}
        0 & (c_f < F) \\
        (1-p)^{c_p}P(a|l)P(l) & (c_f = F)
    \end{cases}
    \label{eq:taper_formula}
\end{equation}

\noindent We later use $\alpha=-\text{log}_2(1-p)$ to control how much to 
weigh the dynamic information: when $\alpha$ is large, $c_p$ will have 
significant sway on the ranking results, while when $\alpha$ is small, $c_p$ 
will have less influence. The impact of $\alpha$ corresponds to the strength of the assumption 
of the statistical model in \Cref{eq:taper_cp_assump}. Thus, by inspecting 
whether test behavior would change (locally) when a patch is applied, we can 
obtain a more precise posterior probability regarding which patch is likely to 
be correct. This technique may also be used to obtain more precise
fault localization results: we may simply marginalize over the space of repair 
actions as follows:
\begin{equation}
    P(l|D) = \sum_a{P((l, a)=\text{fix}|D)}
    \label{eq:fl_formula}
\end{equation}

Based on these derivations, we describe how program states may be efficiently evaluated for this
technique to be practical, and about the specific choices of $P(l)$ and $P(a|l)$.

\subsection{Implementation Overview}

\begin{figure}[h!]
    \centering
    \includegraphics[width=1.0\linewidth]{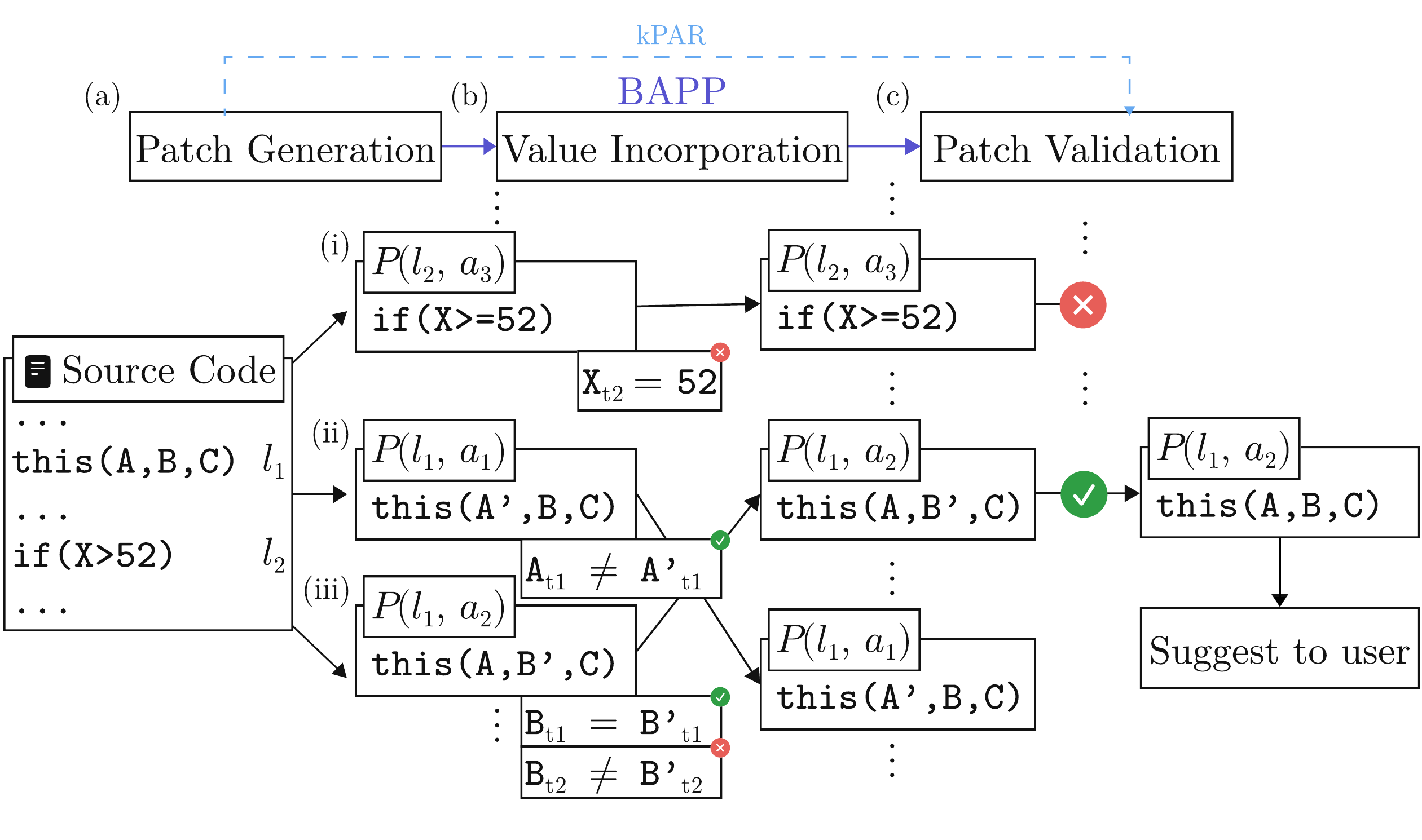}
    \caption{Overview of \name.}
    \label{fig:diagram}
\end{figure}

In the rest of this section, we will explain the implementation details of \name, a patch reranking tool 
built upon the derivations presented in the previous subsection. As shown in \Cref{fig:diagram}, \name
can be broadly divided into three steps. First is the generation of all possible patches, the implementation
of which is closely based on kPAR~\cite{Liu2019kpar}, the open-source implementation of the original
pattern-based APR, PAR~\cite{Kim:2013ty}. In the second step, we use the Java debugger, JDB, to extract values of expressions
relevant to the patches generated in the previous step. This step is our main
contribution to the overall technique, as the original kPAR simply comprises of the first and the third step.
As will be discussed later in the section, this stage also involves the removal of patches with syntax
errors saving the compilation cost from the original kPAR implementation. Using the extracted values, a
likelihood score is calculated for each possible patch in accordance with the derivations presented in the
previous subsection, and these scores are used to comprehensively rank the patches for the next and
final step: patch validation.

\begin{lstlisting}[basicstyle=\footnotesize\ttfamily,
    breaklines=true,
    caption={Abbreviated patch for Chart-8.},
    label={lst:chart8p},]
 public Week (Date time, TimeZone zone) {
-    this(time, RTP.DEFAULT_TIME_ZONE, ...);
+    this(time, zone, ...);
 }
\end{lstlisting}

The following subsections provide further details for each of the steps: patch generation,
value extraction, and validation. In order to provide a clear picture of the entire process, we will use the
correctly generated patch for Defects4J Chart-8 as a running example. The context of the
buggy line in the source code is presented in Listing~\ref{lst:chart8p}. This simple patch is shown 
abbreviated in \Cref{fig:diagram} (iii), along with alternative patches in the same project. 

\subsection{Patch Generation}
\label{sec:approach_patchgen}

\begin{table*}[]
    \scalebox{1.0}{
    \begin{tabular}{|l|l|}
    \hline
    \textbf{Template Type} & \textbf{Description}                                                                         \\ \hline
    Parameter Replacer     & Replace an argument with another variable of the appropriate type. \\ \hline
    Parameter Adder        & Switch to an overloaded method by adding a variable of the appropriate type as an additional argument.   \\ \hline
    Parameter Remover      & Switch to an overloaded method by removing an existing argument.                                         \\ \hline
    Method Replacer        & Replace the method name to another method of the same type from the same class.                          \\ \hline
    Conditional Replacer   & Replace a conditional expression with another boolean expression.                                        \\ \hline
    Conditional Adder      & Append a new component to a conditional expression using \texttt{||} or \texttt{\&\&}.                                     \\ \hline
    Conditional Remover    & Remove a component of a conditional expression.                                                          \\ \hline
    Null Checker           & Insert a null checker before a referenced variable.                                                      \\ \hline
    Cast Checker           & Insert a type checker before a typecasted variable.                                                      \\ \hline
    \end{tabular}
    }
\end{table*}

Our patch generation shares repair templates with kPAR. We 
first generate the AST of all the files covered by the failing tests using the 
\texttt{javalang} library \cite{c2nes2022javalang}. Then, using this AST, we 
find matching templates for each of the lines executed by the failing tests. 
The template types and the possible patches that can be generated for each of 
the templates are presented in Listing~\ref{lst:chart8p}. Looking at the buggy line in our 
example code in Listing~\ref{lst:chart8p}, \name would detect a method invocation node in 
the AST at this location. Traversing the AST also allows us to detect the 
\texttt{zone} variable that can be used to replace the second argument in this 
method invocation. Thus, \name would be able to conclude that a Parameter 
Replacer template could be applied to this location.

After the AST analysis, \name generates all possible patches for each variant 
of the identified matching templates. In the case of our example, in accordance 
with the description for the Parameter Replacer template, a patch will be generated in
which the original argument \texttt{RegularTimePeriod.DEFAULT\_TIME\_ZONE} 
would be replaced with the variable in the scope with the appropriate type, 
\texttt{zone}. Considering a field of type \texttt{Date} declared in this class
(not shown in Listing~\ref{lst:chart8p}), another Parameter Replacer patch could
be generated as shown in \Cref{fig:diagram} (ii).
The output of this stage is the list of all possible patches for each of the locations under consideration.

\subsection{Value Incorporation}
\label{sec:approach_value}

This stage is where \name deviates from kPAR. For kPAR, the patches generated in 
the previous stage are simply relayed to the validation stage, in which the 
patches are applied and evaluated in the order of the SBFL ranking of their
locations. Instead, \name first uses JDB to execute failing and passing tests 
on the original unchanged source code and extracts the values of the original 
expression and the new expression of each patch generated in the 
previous stage. Considering the patch $a_2$ in \Cref{fig:diagram} (iii),
whenever the breakpoint is triggered at location $l_1$, we would extract the
value of the original argument \texttt{RegularTimePeriod.DEFAULT\_TIME\_ZONE},
as well as the value of the new argument \texttt{zone}, illustrated under 
\Cref{fig:diagram} (b) as \texttt{B} and \texttt{B'}, respectively. After the
execution of each test, \name analyzes these values to either filter out
implausible patches or assign a likelihood score 
for the remaining patches. \Cref{eq:taper_formula} derived in the previous 
section summarizes how the values extracted are processed: note that values 
from failing tests are processed differently from values of the passing tests.

All failing tests are executed before any of the passing tests are executed. In 
accordance with the assumption that the fix must change the behavior of the 
failing test, as specified in \Cref{eq:taper_formula}, the value of the original
expression is compared with the value of the new expression for each 
patch. Any patch for which the two values are identical is discarded as 
implausible. In our running example, if \texttt{RegularTimePeriod.DEFAULT\_TIME\_ZONE}
and \texttt{zone} have equal values whenever this particular 
line is executed in a failing test, then this particular patch would be removed 
from the pool of possible patches after the execution of that failing test.

After all failing tests are executed, passing tests are run in order to assign 
a likelihood score to each of the remaining patches after the implausible 
patches have been filtered out. In our implementation, the $c_p$ term in
\Cref{eq:taper_formula} is set to the number of passing tests in 
which the original value and the replacement value are different at every instance 
in which the location in question is executed during the passing test. More 
intuitively, if the patch does not change the behavior of the passing test, the 
likelihood score increases, and vice versa. With the incorporation of normalized 
Ochiai SBFL scores represented in \Cref{eq:taper_formula} as $P(l)$, the final 
score is calculated after the execution of passing tests in accordance with the 
equation.

Although argument values are evaluated for Parameter Replacement patches as 
shown in our example, the return values of the method invocations are not 
evaluated. We empirically find that invoking methods
for value extraction often leads to various side effects, threatening the
integrity of value extraction in other patches and thus the accuracy of
the tool.
For similar reasons, return values are not evaluated 
for Parameter Adder, Parameter Remover, and Method Replacer.

To improve efficiency, we apply the following optimization to this stage. 
First, we only consider the top 200 locations in the SBFL ranking. To prevent 
JDB stopping at breakpoints within loops at every iteration, we limit each 
breakpoint to 100 hits, before which the corresponding values are not extracted. 
Values are only extracted for the \emph{last} 100 hits of a statement,\footnote{statement execution counts can be retrieved from coverage profilers, 
which are used by the SBFL technique.} based on prior work showing that failing values
that induce test failures appear in shorter execution traces~\cite{Assi2019CCDFJ}. We also impose a 15-minute timeout to 
the value-extraction stage, which we found to be reasonable across all bugs we 
studied. With the timeout of 15 minutes, \name often cannot execute all passing 
tests, especially for projects like Closure which has a large number of test 
cases. To address this issue, we prioritize passing test execution based on the 
current likelihood score of all the lines covered by each remaining passing 
test.

\subsection{Patch Validation}
\label{sec:approach_validation}

Through the previous steps, we have generated all the possible patches -- which 
essentially means that all the information necessary to apply the patches have 
been collected -- and these patches have been ranked based on the relevant 
information including the program states and the Ochiai SBFL results. The final 
application and evaluation of the patches in the specified order is performed 
by replacing the original expression in the source code with the new expression.

For each patch in the ranking, we first apply the patch to the source code. All 
the failing tests are run before the passing tests, and during the runs, if any 
of the tests fail, the patch in question is considered as faulty and the next 
patch is considered. When a patch is found which passes all the failing and 
passing tests, the repair process is terminated, as illustrated under 
\Cref{fig:diagram}(c). It is important to note that 
some bugs may have multiple plausible patches. Because \name simply terminates 
after finding the first plausible patch, an incorrect plausible patch 
might be output instead of the correct patch.

\section{Experimental Setup}
\label{sec:setup}

This section describes the settings of our empirical studies.

\subsection{Configurations}
\label{sec:configurations}

As mentioned in the previous sections, we use \texttt{javalang}~\cite{
c2nes2022javalang} for the generation of AST for the source code, and JDB for 
the extraction of values of relevant expressions during test executions. During 
the implementation of \name, we encountered inconveniences that motivated us to 
make changes to the \texttt{javalang} and JDB modules, in order to fix bugs or add 
features. For instance, we added a feature to convert a part of the AST tree 
back into code, which was not originally provided in \texttt{javalang}. Other 
changes include adding the position information to node types for which the 
information was originally omitted. When using the JDB, it was necessary to 
make changes to the module in order to ensure that JDB has the same execution 
semantics as the native Java runtime: for example, JDB originally lacks support 
for short-circuit evaluation.

Although \name's patch generation was based on kPAR's implementation, our 
results have several differences with kPAR's results that are worth noting. 
First, while kPAR's results are based on Defects4J v1, some bugs of which have been 
modified for Defects4J v2. For the sake of our experiment, we have  
excluded bugs that kPAR was able to patch with multi-hunk patches. Finally,
kPAR uses information about methods defined in external modules in order to generate 
patches for templates such as Method Replacers and Parameter Replacers. Due to
the limitations of \texttt{javalang}, we omit support for patching invocations to 
external methods. 

We used Ochiai \cite{ochiai1957zoogeographic} suspiciousness order for our SBFL 
ranking ($P(l)$), and use the uniform distribution for $P(a|l)$; that is,
if the number of patches generated by kPAR at a location is $N_l$, $P(a|l)=\frac{1}{N_l}$. 
This causes our ranking to be different from that of kPAR even when there is no dynamic information.
Ties in both SBFL and our FL technique were broken using the max-rank
tiebreaker, as is done in prior FL research~\cite{Sohn2017FLUCCS}. To evaluate FL
results, we use the acc@$k$ metric, which evaluates how many bugs can be localized
within $k$ inspections.
The $\alpha$ parameter was set to 3 in RQ1 and RQ2 as it empirically
showed the best performance. The experiments were run on machines with Intel(R) Core(TM) i7-6700 
CPU @ 3.40GHz and 32GB of DDR4 RAM @ 2133MHz.

\subsection{Research Questions}
\label{sec:research_questions}

We aim to answer the following research questions with our empirical evaluation.

\noindent\textbf{RQ1. Efficiency Improvement:} How much more efficient is \name in comparison to kPAR in finding the
first plausible patch? For this question we consider execution time and the overall patch rank of the first
plausible patch.

\noindent\textbf{RQ2. FL Improvement:} How much improvement can be made to the SBFL ranking of the true buggy line by
incorporating the likelihood score calculated from the value extraction?

\noindent\textbf{RQ3. Configuration Study:} How much weight should be given to program states and SBFL for patch ranking?
For this question we consider different values of $\alpha$ in \Cref{eq:taper_formula}. In addition, we evaluate
patch rankings prioritizing SBFL results while using program states only as a tiebreaker, and vice versa.

\noindent\textbf{RQ4. Qualitative Analysis:} When does \name perform well, and when does it not?
We analyze the reasons behind the successes and failures of \name, providing a breakdown of cases.

\section{Results}
\label{sec:results}

This section presents the results from our empirical evaluation.

\subsection{RQ1: Efficiency Improvement}
\label{sec:rq1}

Out of 41 Defects4J bugs successfully patched by kPAR with our experimental setup, all 41 bugs are
successfully patched with \name. This indicates that the patches filtered out during the value incorporation
stage does not include plausible patches necessary to fix the bugs. As mentioned in \Cref{sec:approach_value},
the patches that do not change program behavior during the execution of failing tests are filtered out, as
well as patches that are predicted to cause compilation error if applied to the source code.

\begin{figure}[h!]
    \centering
    \subfigure[Raw Ranks.]{\includegraphics[width=0.49\linewidth]{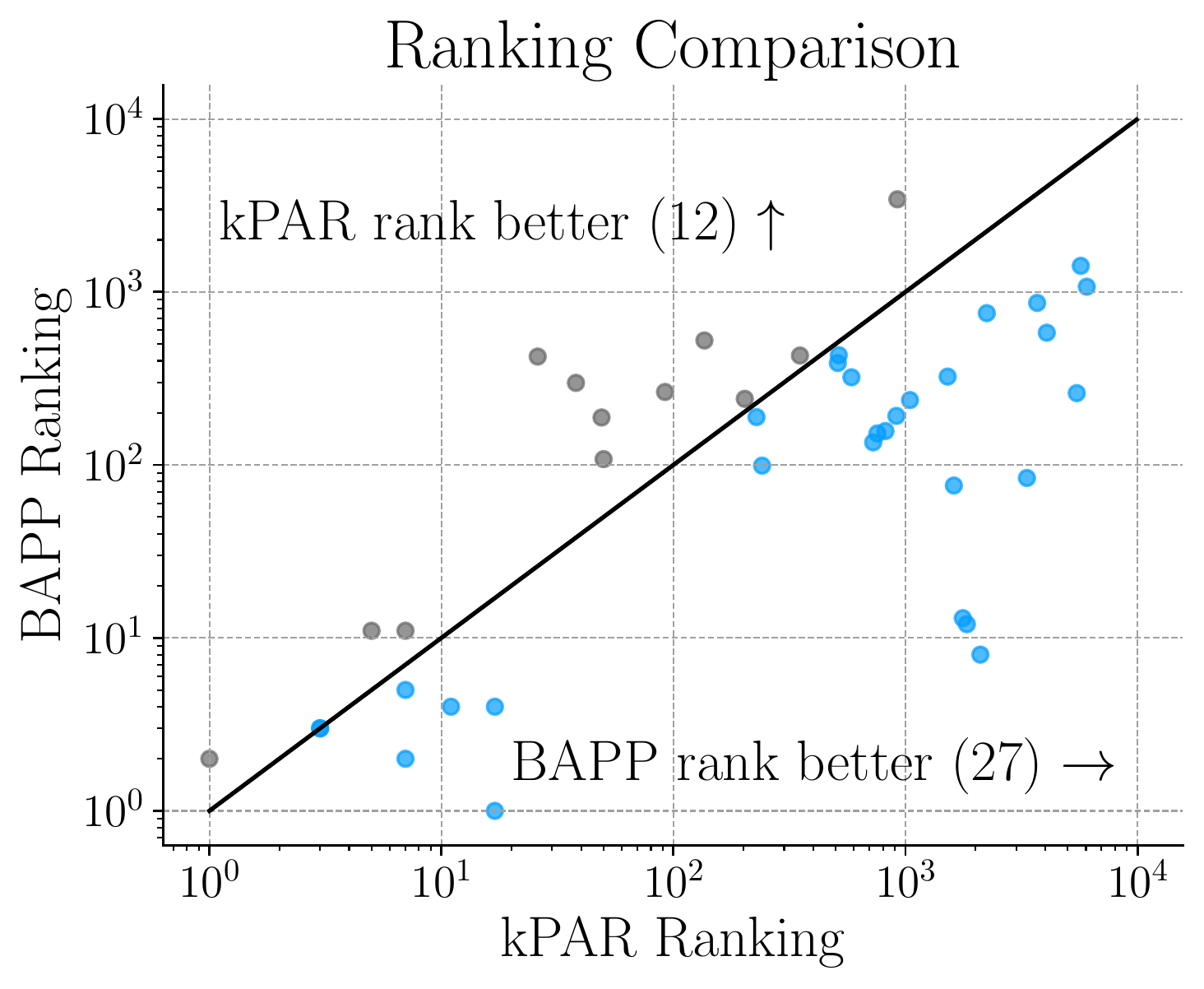}}
    \subfigure[Rank Ratios.]{\includegraphics[width=0.49\linewidth]{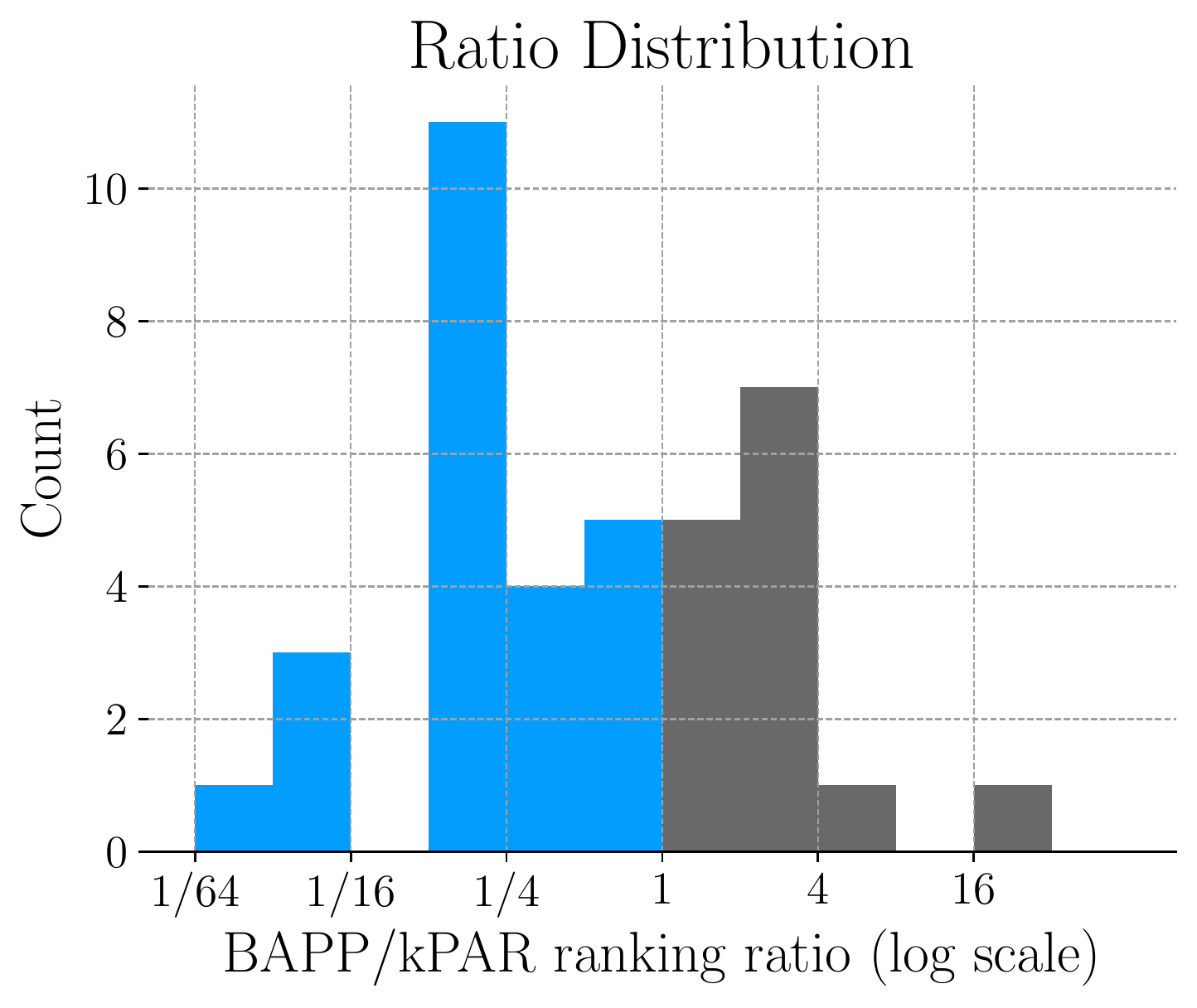}}
    \caption{Performance of \name at reranking patches.}
    \label{fig:rq1}
\end{figure}

The improvement in the efficiency of kPAR's APR process with the introduction of \name is shown in
\Cref{fig:rq1}. \Cref{fig:rq1}(a) plots kPAR and \name's ranks of the first 
plausible patch that was evaluated, illustrating the difference in efficiency 
for bugs from projects of different sizes. Overall, we observe consistent 
improvements across bugs of all project sizes. This is noteworthy because bugs 
from large projects also include a large number of tests that need to be run 
during the value-incorporation stage. However, because of the 15-minute 
timeout set on the execution of this stage as described in 
\Cref{sec:approach_value}, only tens or hundreds of tests out of thousands can 
be executed to extract the program states. The fact that these bugs saw
significant improvements in efficiency indicates the effectiveness of the 
optimization described at the end of \Cref{sec:approach_value}.

Out of 41 bugs studied, 13 are related to method invocation (i.e. Parameter 
Replacer/Adder/Remover and Method Replacer). As explained in 
\Cref{sec:approach_value}, we do not extract return values for lines that fit 
these templates, because of the side effects from the duplicate method 
invocation necessary to extract these values. While one may
wonder if our technique will also show improvements in those cases, 
many of these patches in fact show significant improvements in the rank of the 
plausible patch. In fact, one of the biggest improvements that can be seen in 
\Cref{fig:rq1}(a) is for Closure-10, whose patch is of the Method Replacer
type; \name improves the rank of the plausible patch to 83,
from kPAR's 3338. From such examples, we infer that even if values cannot be 
extracted for the plausible patch, the value extraction and processing for the rest of the patches can yield a ranking that ultimately improves the efficiency for many bugs.

\Cref{fig:rq1}(b) plots the \name/kPAR ratio for the ranks of plausible 
patches: the peak of the distribution is between $\frac{1}{8}$ and 
$\frac{1}{4}$, indicating that the efficiency improvement with respect to 
patch ranking for \name is within four- to eight-fold for a large portion of the bugs under consideration. The median ratio is 0.32.


So far, we have compared the efficiency of \name with respect to kPAR in terms 
of the rank of the plausible patch. However, execution time comparison gives a 
better picture of the practical effectiveness of \name. This is because, as 
mentioned in \Cref{sec:approach_validation}, all patches suspected of
causing compilation errors are filtered out from the pool of patches to be 
evaluated. On the other hand, the presence of patches with compilation errors 
means that kPAR takes less time than average to evaluate a patch. The mean 
execution-time reduction across the 41 bugs is 34 minutes, despite the 
overhead of value extraction which times out at 15 minutes. Thus, we argue that 
the improvement in efficiency outweighed the overhead cost of the extraction 
and evaluation of program states.

\begin{tcolorbox}[boxrule=0pt,frame hidden,sharp corners,enhanced,borderline north={1pt}{0pt}{black},borderline south={1pt}{0pt}{black},boxsep=2pt,left=2pt,right=2pt,top=2.5pt,bottom=2pt]
    \textbf{Answer to RQ1:} With the \name/kPAR ratio of the ranking of the first plausible patch at
    a median of 0.32, the incorporation of program states introduced improvements in efficiency for
    bugs from small projects as well as bugs from large projects. The mean execution-time reduction of 34 minutes 
    demonstrates the practical efficiency improvement
    outweighed the overhead cost of value incorporation.

\end{tcolorbox}

\begin{figure}[h!]
    \centering
    \subfigure[Raw Ranks.]{\includegraphics[width=0.49\linewidth]{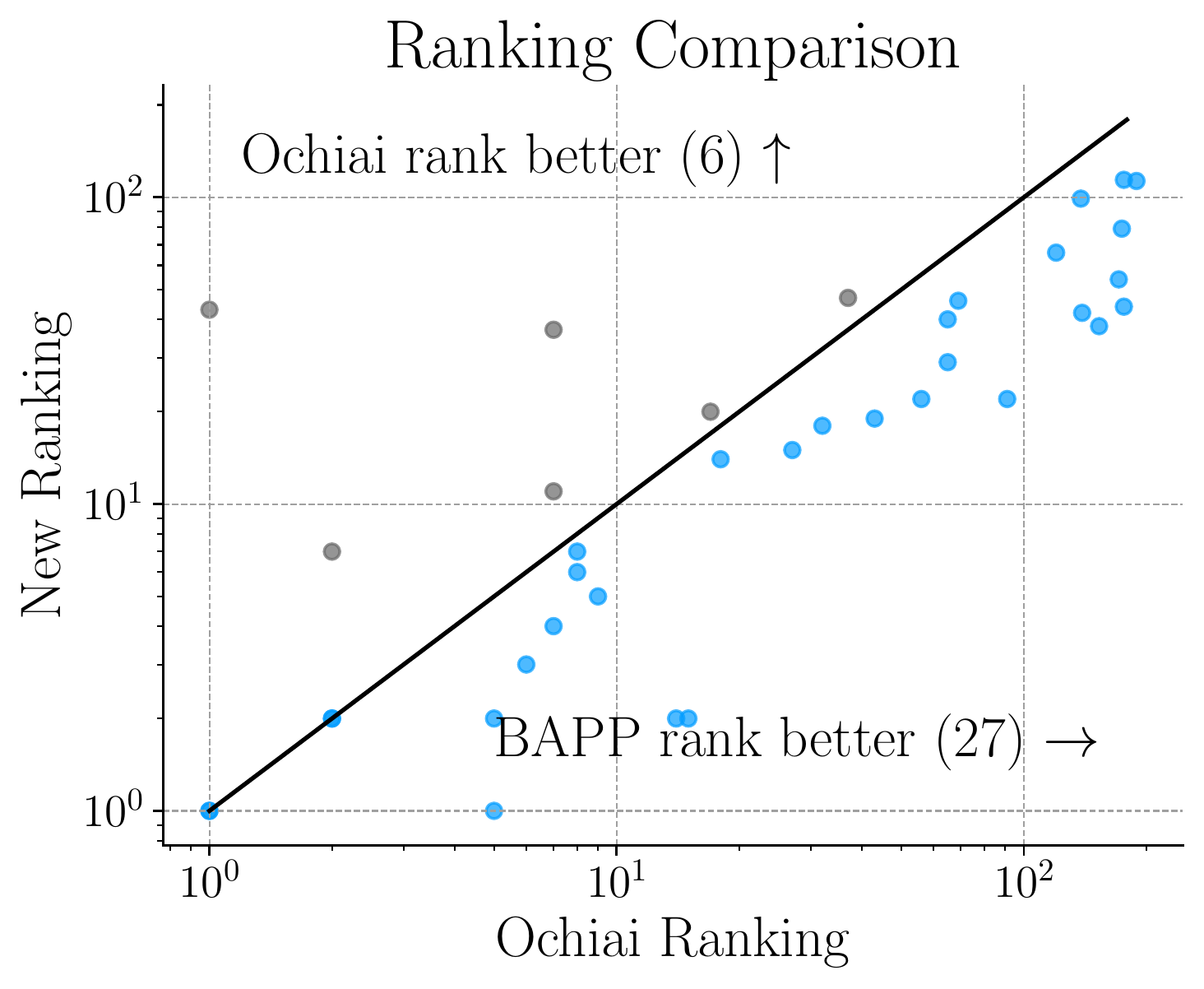}}
    \subfigure[acc@$k$.]{\includegraphics[width=0.49\linewidth]{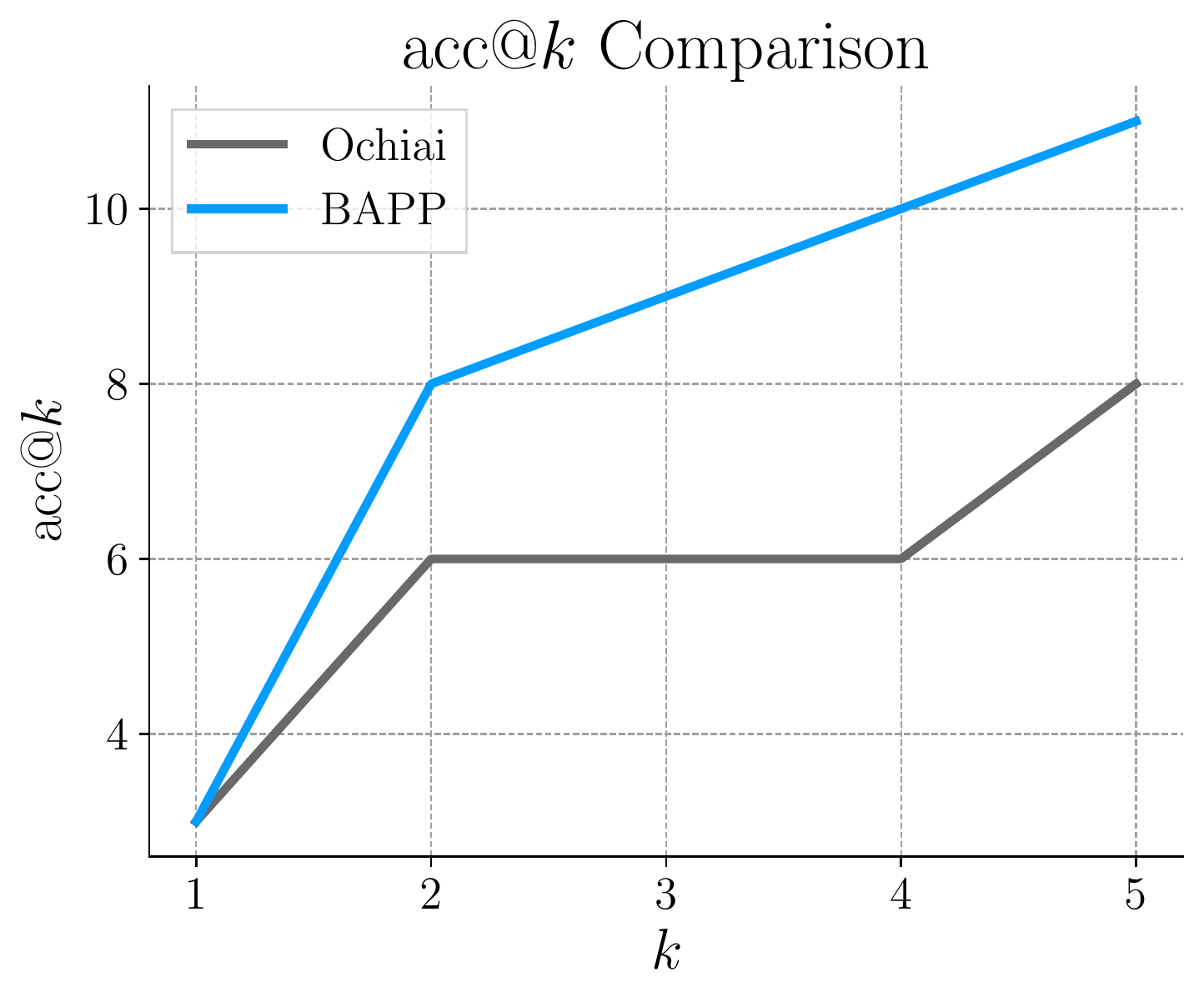}}
    \caption{Performance of \name at FL.}
    \label{fig:rq2}
\end{figure}

\subsection{RQ2: FL Improvement}
\label{sec:rq2}

In accordance with \Cref{eq:fl_formula}, we built a new FL ranking of the 
covered locations for each of the Defects4J bugs under consideration. 
\Cref{fig:rq2}(a) shows the changes in the ranks of the true buggy line 
between SBFL and \name's FL across bugs from projects of different sizes. The 
differences in the FL rankings resemble the differences in patch rankings 
shown in \Cref{fig:rq1}. The median of the \name-FL/SBFL ranking ratio is 
0.57, indicating general improvements.
Additionally, \Cref{fig:rq2}(b) shows the acc@k comparison between SBFL and \name-FL: we 
find that \name-FL generally outperforms SBFL. Existing work has shown that 
identifying the true fault location within a few tries is important for the 
developer trust in FL techniques~\cite{Kochhar2016FLExpectation}. We believe 
these results indicate that \name shows promise in improving practical fault 
localization as well, with a relatively small computational budget of at most 
15 minutes.

\begin{tcolorbox}[boxrule=0pt,frame hidden,sharp corners,enhanced,borderline north={1pt}{0pt}{black},borderline south={1pt}{0pt}{black},boxsep=2pt,left=2pt,right=2pt,top=2.5pt,bottom=2pt]
    \textbf{Answer to RQ2:} 
    \name's FL performed better than Ochiai overall, with the reduction in rank of the true buggy line at a median of
    0.43, and identified the true buggy location at top-k rankings more often as well.
\end{tcolorbox}

\subsection{RQ3: Configuration Study}

\begin{figure}[h!]
    \centering
    \includegraphics[width=\linewidth]{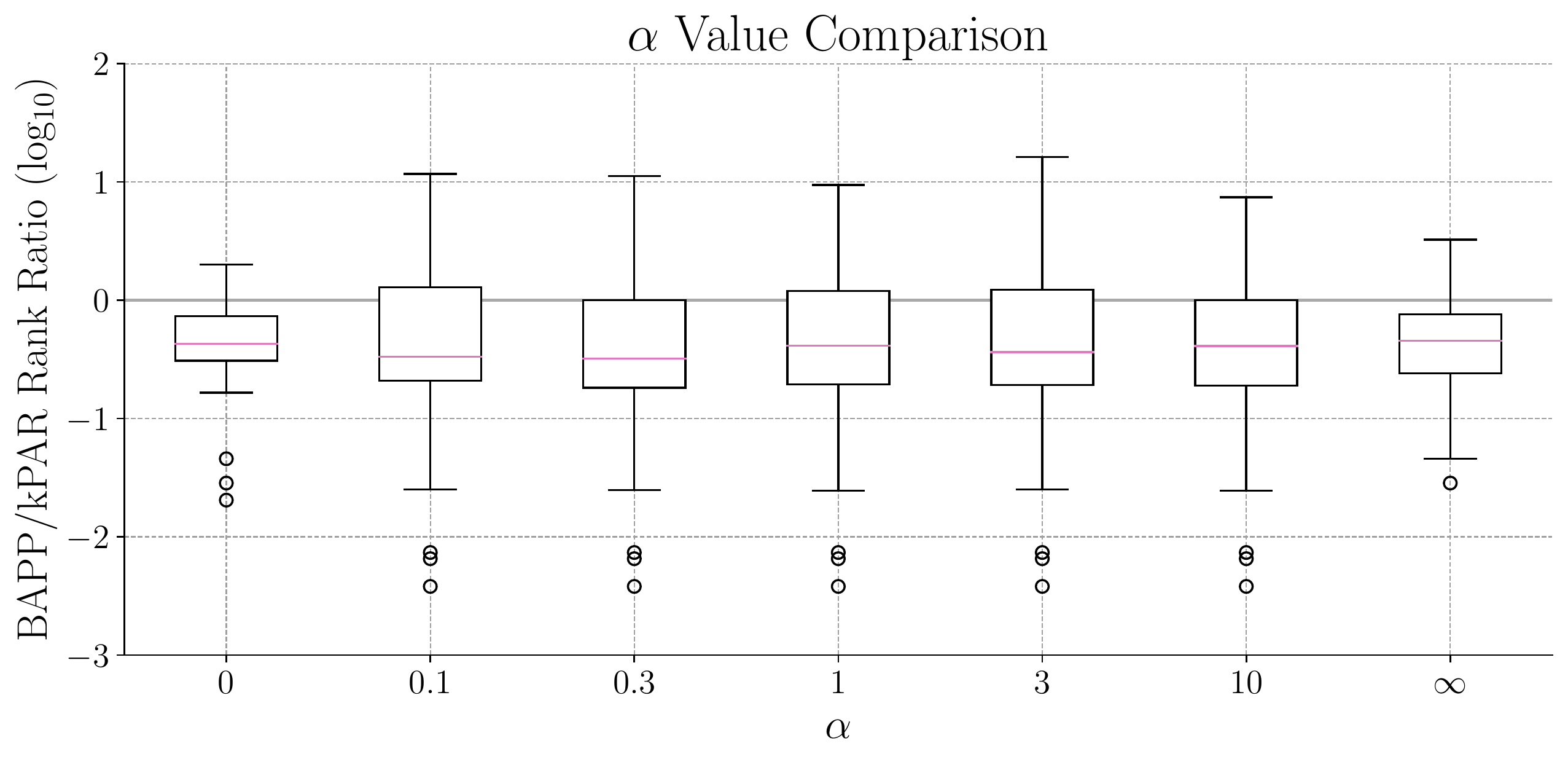}
    \caption{Performance of \name under different $\alpha$.}
    \label{fig:rq3}
\end{figure}

\Cref{fig:rq3} depicts the patch ranking reduction ratio of \name as $\alpha$ changes;
among the five values of $\alpha$ for which \name was evaluated, $\alpha=0.3$ saw the
greatest reduction in the plausible patch ranking, as shown in \Cref{fig:rq3}(a). 
While not in the figure, we found that FL performance was best when $\alpha=3$ as well.
Nonetheless, the result distributions over different $\alpha$ values shows little difference,
indicating the performance is resilient to specific values of $\alpha$. Thus, in general
the incorporation of program values is enough to enhance APR and FL performance.

When the scores in which the $c_p$ term was prioritized in ranking the patches while the
$P(l)$ term was simply used as a tiebreaker for patches with identical (essentially $\alpha$ value
of infinity), the median reduction ratio of rank was 0.45. Compared with the median reduction ratio
of 0.32 when $\alpha=3$, this demonstrates the practical effectiveness of considering
both the $c_p$ term and the $P(l)$ term when ranking the patches, as derived in
\Cref{sec:approach_derivation}. Further, when compared to completely prioritizing fault localization
$\alpha=0$, we find that multiplying FL and dynamic information shows superior performance by
31\%, supporting our earlier point that there may be a better way to use patch probabilities
generated by G\&V techniques.

\begin{tcolorbox}[boxrule=0pt,frame hidden,sharp corners,enhanced,borderline north={1pt}{0pt}{black},borderline south={1pt}{0pt}{black},boxsep=2pt,left=2pt,right=2pt,top=2.5pt,bottom=2pt]
    \textbf{Answer to RQ3:} The $\alpha$ value of 1 yielded the highest efficiency for identifying
    plausible patches, although different $\alpha$ values tested had similar results.
\end{tcolorbox}

\subsection{RQ4: Qualitative Analysis}

We first present a breakdown of the individual cases in which \name underperformed its counterpart
technique. First, when performing patch ranking, there were two main reasons plausible patches
were ranked lower than the initial ranking from kPAR. In some cases, the statistical model that
we used did not favor the patch: certain plausible patches would change passing test behavior often,
or even whenever they were executed. For example, we found that the state would always change for
the Chart-8 bug; nonetheless the patch itself is correct. In other cases, due to the large number
of patches generated at certain locations, the likelihood of patches at those locations would drop
due to the $P(a|l)=\frac{1}{N_l}$ term. As a result, all patches from such locations would be
deprioritized, leading to worse results. For example, a plausible patch for the Math-15 bug shared
patch location with 205 other patches, and as a result dropped in ranking. We are exploring better
formulations of the task that would not suffer from this issue.

Our analysis for FL similarly reveals two reasons our technique yielded worse results. One issue was that due
to the nature of our FL technique which is closely related to patch templates, our technique could
not suggest statements for which no patch was generated. For example, in Closure-22, one of the
top-ranked actual buggy locations is simply a \texttt{continue;} statement, for which our technique generates
no patches, and consequently fails to rank. We believe such issues can be overcome by adopting
more flexible patch generation techniques in future work. A second issue was that for certain
statements with conditions, a large number of patches that would always change the state would be
generated, and as a result the likelihood of the statement (which is the sum of the likelihood
of patches) would drop.

On the other hand, when such pitfalls are not met, our technique performs well; 
in Math-85, for example, the correct patch replaces the conditional expression 
\texttt{fa * fb >= 0.0} with \texttt{fa * fb > 0.0}. For every instance in which this 
buggy line was hit during failing test executions, \texttt{fa * fb} was equal to 
\texttt{0.0}, meaning the patch would change the program behavior. On the other hand, 
\texttt{fa * fb} was never equal to \texttt{0.0} during the execution of any of the passing 
tests in this project, leading \name to improve the FL ranking by 76\% ($91 \rightarrow 22$)
and the APR ranking by 81\% ($820 \rightarrow 159$).

\begin{tcolorbox}[boxrule=0pt,frame hidden,sharp corners,enhanced,borderline north={1pt}{0pt}{black},borderline south={1pt}{0pt}{black},boxsep=2pt,left=2pt,right=2pt,top=2.5pt,bottom=2pt]
    \textbf{Answer to RQ4:} We broadly identify issues hindering better performance of \name, such
    as patches that do not closely match the statistical model we use.
\end{tcolorbox}

\subsection{Threats to Validity}
\label{sec:threats}

Threats to \textbf{internal validity} concern whether the results presented in the paper are sound.
We believe there is little concern regarding the Bayesian framework we proposed.
In the case of \name,
we take account of the potential idiosyncrasy of different bugs by experimenting over 41 bugs from
the widely-used Defects4J benchmark of real-world faults. Further, we perform a search over
the parameter $\alpha$ in RQ3, showing that performance gradually changes as the parameter changes.

Threats to \textbf{external validity} concern whether the results would generalize to new subjects.
We have attempted formulating techniques from a broad cross-section of the automated debugging
literature using our proposed theoretical framework; as long as a technique shares the goal of
inferring the posterior likelihood of the correct patch $P(l, a)$ we believe our framework
will continue to be applicable. Meanwhile, we have presented results of \name reranking patches
generated by kPAR; while our analysis shows that our simple statistical model works well for
kPAR-generated patches, further experimentation is needed to decide whether our assumptions work for other patch generation techniques.

\section{Discussions \& Future Work}
\label{sec:discussion}

A major limitation of our framework is the single fault assumption, limiting 
the cases to which our theories can be applied. While it is possible to 
overcome these issues by reasoning over \emph{sets} of solutions instead of 
single solutions as we have done in our work, when there are $N$ possible 
solutions this requires reasoning over $O(2^N)$ combinations of solutions, 
which quickly becomes impractical. Barinel~\cite{Abreu:2009qy} uses a 
heuristic named Staccato~\cite{Abreu2009Staccato} to generate a smaller group 
of candidates to perform Bayesian inference over; more experiments are 
required to determine whether such heuristics would be scalable for automated 
debugging in general, and not just fault localization.

Our framework also directs us towards future research directions that we hope 
to pursue further. To start off, we consider how existing techniques deal with 
the prior probability of patches, $P(l, a)$. While in almost all APR work it 
is decomposed to $P(a|l)P(l)$ and thus fault localization precedes patch 
generation, it does not necessarily need to be this way. Under our framework, 
one can equally decompose $P(l, a)$ to $P(l|a)P(a)$ instead, identifying the 
repair operation prior to performing fault localization. In certain cases, this 
formulation is closer to human practice: for example, in Defects4J Lang-29, 
the error message shows `expected: [0] but was: [0.0]', from which one can 
infer that (i) a type needs to be changed somewhere, but (ii) which location 
to fix is unknown. Indeed, some existing techniques have actually pioneered 
this concept in a restricted way: VFix~\cite{Xu2019VFix} notably focuses on 
null pointer exception fixes, and searches for fix locations given the types 
of fixes it can do. Such a direction is particularly promising given the 
recent improvements in using error messages for generating 
patches~\cite{Ye2022SelfAPR}.

Finally, while the automated debugging work covered in this paper do not 
incorporate dependency information and updates at most based on the evaluation 
results of patches generated at the same location, we believe our model could 
facilitate the derivation of FL and APR technologies that leverage dependency 
information such as call stacks, and thus enhance precision; we hope to pursue 
such research areas in future work.

\section{Conclusion}
\label{sec:conclusion}

We propose a Bayesian framework of automated debugging, postulating that the 
ultimate goal of automated debugging techniques is to infer the posterior 
likelihood over the space of fault locations and repair actions, $P(l, a)$. We 
find that this formulation can recover previously proven results, such as the 
maximality of the Op2/Binary SBFL formulae, as well as have specific
probability terms neatly mapped to specific automated debugging concepts and 
allow an inspection of the assumptions behind automated debugging techniques. 
To demonstrate the utility of the framework, we propose a novel 
value-incorporating patch prioritization technique for APR, whose core 
principles are derived from our Bayesian framework. Along with the use of 
debuggers which allows the efficient implementation of the recommendations of 
the framework, we find that overall our tool \name can improve the patch 
ranking by 68\%, leading to an average execution time reduction of 34 minutes. 
\name also improves the FL ranking in two-thirds of the inspected bugs, 
leading to an increase in acc@k values. In addition, our ablation study 
reveals \name is resilient to the choice of $\alpha$ values. We believe that 
our Bayesian framework also suggests interesting research directions that 
have not been thoroughly explored and hope to perform related research in the 
future.

\section{Acknowledgement}
This research was supported by the KAIST-Samsung SDS joint research center through the Project of Code Representation Learning
and the Undergraduate Research Project programme at KAIST.



\bibliographystyle{IEEEtran}
\bibliography{newref}

\end{document}